%% file: manuscript-arXiv-20111213.tex
\documentclass[twocolumn,prb,showpacs,floatfix,superscriptaddress]{revtex4-1}

\usepackage{amsmath, amsfonts, amssymb, amsbsy}
\usepackage{graphicx}
\usepackage{xspace}
\usepackage{tabularx}

\newcolumntype{C}[1]{>{\centering\arraybackslash}p{#1}}

\newcommand{\lem}{LE-$\mu$SR\xspace}

\newcommand{\trimsp}{\texttt{TRIM.SP}\xspace}
\newcommand{\PBCO}{PrBa$_{2}$Cu$_{3}$O$_{7-\delta}$\xspace}
\newcommand{\YBCO}{YBa$_{2}$Cu$_{3}$O$_{7-\delta}$\xspace}
\newcommand{\YPBCO}{Y$_{1-x}$Pr$_{x}$Ba$_{2}$Cu$_{3}$O$_{7-\delta}$\xspace}
\newcommand{\STO}{SrTiO$_{3}$\xspace}
\newcommand{\Tc}{$T_{\mathrm{c}}$\xspace}
\newcommand{\eg}{\emph{e.\,g.}\xspace}

\begin{document}
\title{Magnetism, superconductivity and coupling in cuprate heterostructures probed by low-energy muon-spin rotation}

\author{B.~M. Wojek}
\altaffiliation{Present address: KTH Royal Institute of Technology, ICT Material Physics, Electrum 229, 164 40 Kista, Sweden}
\email{basti@kth.se}
\affiliation{Labor f{\"u}r Myonspinspektroskopie, Paul Scherrer Institut, 5232 Villigen PSI, Switzerland}
\affiliation{Physik-Institut der Universit{\"a}t Z{\"u}rich, Winterthurerstrasse 190, 8057 Z{\"u}rich, Switzerland}
\author{E.~Morenzoni}
\email{elvezio.morenzoni@psi.ch}
\affiliation{Labor f{\"u}r Myonspinspektroskopie, Paul Scherrer Institut, 5232 Villigen PSI, Switzerland}
\author{D.~G. Eshchenko}
\altaffiliation{Present address: Bruker BioSpin AG, Industriestrasse 26, 8117 F\"allanden, Switzerland}
\affiliation{Labor f{\"u}r Myonspinspektroskopie, Paul Scherrer Institut, 5232 Villigen PSI, Switzerland}
\affiliation{Physik-Institut der Universit{\"a}t Z{\"u}rich, Winterthurerstrasse 190, 8057 Z{\"u}rich, Switzerland}
\author{A.~Suter}
\affiliation{Labor f{\"u}r Myonspinspektroskopie, Paul Scherrer Institut, 5232 Villigen PSI, Switzerland}
\author{T.~Prokscha}
\affiliation{Labor f{\"u}r Myonspinspektroskopie, Paul Scherrer Institut, 5232 Villigen PSI, Switzerland}
\author{H.~Keller}
\affiliation{Physik-Institut der Universit{\"a}t Z{\"u}rich, Winterthurerstrasse 190, 8057 Z{\"u}rich, Switzerland}
\author{E.~Koller}
\affiliation{D\'epartement de Physique de la Mati\`ere Condens\'ee, Universit\'e de Gen\`eve, 1211 Gen\`eve 4, Switzerland}
\author{\O{}.~Fischer}
\affiliation{D\'epartement de Physique de la Mati\`ere Condens\'ee, Universit\'e de Gen\`eve, 1211 Gen\`eve 4, Switzerland}
\author{V.~K. Malik}
\altaffiliation{Present address: Department of Chemical Engineering and Materials Science, University of California Davis, Davis, California 95616, USA}
\affiliation{D\'epartement de Physique, Universit\'e de Fribourg, 1700 Fribourg, Switzerland}
\author{C.~Bernhard}
\affiliation{D\'epartement de Physique, Universit\'e de Fribourg, 1700 Fribourg, Switzerland}
\author{M.~D\"obeli}
\affiliation{Labor f\"ur Ionenstrahlphysik, Eidgen\"ossische Technische Hochschule Z\"urich, 8093 Z\"urich, Switzerland}

\date{\today}

\begin{abstract}
We present a low-energy muon-spin-rotation study of the magnetic and superconducting properties of \YBCO/\PBCO trilayer and bilayer heterostructures. By determining the magnetic-field profiles throughout these structures we show that a finite superfluid density can be induced in otherwise semiconducting \PBCO layers when juxtaposed to \YBCO ``electrodes'' while the intrinsic antiferromagnetic order is unaffected.
\end{abstract}
\maketitle

\section{Introduction}
For more than two decades cuprate thin-film-heterostructure junctions have been studied in order to use them as possible Josephson devices but also to elucidate the transport mechanisms in the various layers. One of the most peculiar barrier materials is \PBCO which depending on the preparation shows a large diversity in its transport properties. A large amount of data has been collected on junctions composed of one high-temperature-superconducting (HTS) \YBCO (YBCO) electrode (and either another YBCO or a Au counter-electrode) and an isostructural but ``semiconducting'' \PBCO (PBCO) barrier layer.\cite{Barner-APL-1991, Gao-JAP-1992, Hashimoto-APL-1992, Boguslavskij-PhysicaC-1992, Golubov-PhysicaC-1994, Satoh-IEEETransApplSupercond-1995, Yoshida-PRB-1997, Kabasawa-PRL-1993, Yoshida-PRB-1996, Suzuki-PRL-1994, Boguslavskij-PhysicaB-1994, Bari-PhysicaC-1996, Faley-IEEETransactApplSupercond-1995, Faley-IEEETransactApplSupercond-1997} The efforts have concentrated on the fabrication and characterization of $a$,$b$-axis oriented junctions. Depending on the preparation, PBCO barriers display a large variety of transport characteristics. On the one hand, metallic layers with normal-state coherence length $\xi_{\mathrm{n}}$ ranging from $5~\mathrm{nm}$ up to $30~\mathrm{nm}$ have been reported.\cite{Barner-APL-1991, Gao-JAP-1992, Hashimoto-APL-1992} On the other hand, insulating layers without\cite{Boguslavskij-PhysicaC-1992} or with contributions of resonant tunneling through a few localized states with a localization length of the order of $1~\mathrm{nm}$ to $3~\mathrm{nm}$\cite{Golubov-PhysicaC-1994, Satoh-IEEETransApplSupercond-1995, Yoshida-PRB-1997} have also been found. The enhanced quasiparticle transport through insulating layers with thicknesses of the order of a few tens of nanometers has been explained by a ``long-range proximity effect'' emerging from resonant tunneling\cite{Devyatov-JETPLett-1994, Devyatov-JETP-1997} or by an alternative scenario in which small CuO-chain-ordered conducting segments serve as negative-$U$ centers with attractive Cooper-pair interaction.\cite{Halbritter-PRB-1992} Eventually, it has been shown that the CuO chains are metallic in structurally ordered PBCO films\cite{Lee-PRB-1996} and that variable-range hopping (VRH) as observed in insulating bulk PBCO\cite{Fisher-PhysicaC-1991} dominates the in-plane transport of oxygen-deficient heterostructures.\cite{Kabasawa-PRL-1993, Yoshida-PRB-1996, Suzuki-PRL-1994}\\
Junctions with $c$-axis orientation have been studied to lesser extent. In this orientation the PBCO barrier layers are mostly found to be insulating and show hopping conduction of quasiparticles through a small number of localized states.\cite{Boguslavskij-PhysicaB-1994, Bari-PhysicaC-1996} However, it has also been argued that an improvement of the film morphology and of the interfaces leads to a metallic behavior with $\xi_{\mathrm{n}}=4~\mathrm{nm}$\cite{Faley-IEEETransactApplSupercond-1995} and that metallicity in PBCO on about the same length scale might be induced due to the proximity to the YBCO electrodes.\cite{Faley-IEEETransactApplSupercond-1997}\\
Superlattice structures consisting of thin layers of superconducting YBCO separated by insulating or ``semiconducting'' layers of either PBCO or \YPBCO have received great attention---\eg for investigations of dimensional effects in ultrathin superconducting layers\cite{Triscone-PRL-1990} or of the interlayer coupling which has been found to be surprisingly large.\cite{Fivat-JLowTempPhys-1996} An extensive review on such studies can be found in Ref.~\onlinecite{Triscone-RepProgPhys-1997}.\\
More recently, HTS Josephson devices regained interest through the reports of the so-called ``giant proximity effect'' and ``giant magneto-oscillations'' in junctions with barriers in the pseudogap\cite{Bozovic-PRL-2004, Morenzoni-NatCommun-2011} or antiferromagnetic (AF) state,\cite{Komissinskiy-PRL-2007} respectively.\\
In this article we report on experiments involving $c$-axis oriented YBCO/PBCO trilayer and bilayer thin-film heterostructures studied by low-energy muon-spin rotation (\lem), a local technique which allows to obtain depth- and layer-resolved information on the superconductive and magnetic properties throughout the structures. We show that layers of PBCO with a thickness of a few tens of nanometers are able to transmit supercurrents when juxtaposed to YBCO layers, while the intrinsic AF order in the {CuO$_2$} planes is hardly disturbed. However, this induced superconductivity is suppressed by the application of a moderate magnetic field.

\section{Sample preparation and characterization}
\begin{figure}
\centering
\includegraphics[width=0.9\columnwidth]{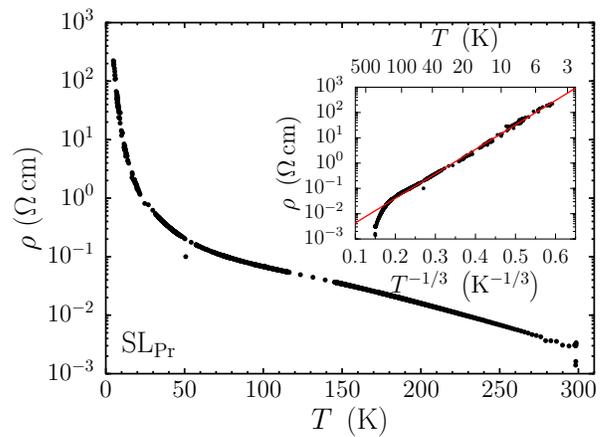}
\caption{(Color online) Temperature dependence of the resistivity of the SL$_{\mathrm{Pr}}$ reference sample. SL$_{\mathrm{Pr}}$ shows semiconducting 2D VRH behavior (measured on a film without {Au} capping layer). The solid red line in the inset is a fit between $5~\mathrm{K}$ and $100~\mathrm{K}$ using the 2D VRH formula given in the text.}
\label{fig:PBCOresistivity}
\end{figure}

\begin{table}
\caption{Thin-film YBCO/PBCO heterostructures studied by \lem. The constituents are indicated starting from the vacuum interface.}
\label{tab:PBCO-studied-films}
\begin{tabularx}{\columnwidth}{C{0.11\columnwidth} C{0.44\columnwidth} C{0.19\columnwidth} C{0.07\columnwidth} C{0.09\columnwidth}}
\noalign{\smallskip}
\hline
\noalign{\smallskip}
thin film & constituents & layer thickness~(nm) & \Tc (K) & growth \tabularnewline
\noalign{\smallskip}
\hline
\noalign{\smallskip}
SL$_{\mathrm{Y}}$ & Au/YBCO & $4$/$200$ & $86$ & MS \tabularnewline
SL$_{\mathrm{Pr}}$ & Au/PBCO & $4$/$45$ & --- & MS \tabularnewline
TL & Au/YBCO/PBCO/YBCO & $4$/$70$/$45$/$75$ & $88$ & MS \tabularnewline
BL$_{\mathrm{Y/Pr}}$ & YBCO/PBCO & $70$/$75$ & $86$ & MS \tabularnewline
BL$_{\mathrm{Pr/Y}}$ & PBCO/YBCO & $70$/$75$ & $88$ & PLD \tabularnewline
\noalign{\smallskip}
\hline
\end{tabularx}
\end{table}
The $c$-axis oriented YBCO/PBCO heterostructures with $\delta = 0.15(5)$ were grown on $10\times 10\times 0.5~\mathrm{mm^3}$ \STO $(100)$ substrates by off-axis rf magnetron sputtering (MS) and by pulsed laser deposition (PLD). Typical parameters for the MS deposition can be found in Ref.~\onlinecite{Schlepuetz-PRB-2010}. The PLD uses a KrF excimer laser with $25~\mathrm{ns}$ pulse length and a central wavelength of $248~\mathrm{nm}$; the fluence on stoichiometric pressed targets is $2.5~\mathrm{J/cm^2}$. During the film growth the substrate was heated to $800~\mathrm{^{\circ}C}$ in an oxygen atmosphere with a pressure of $0.34~\mathrm{mbar}$. After deposition the oxygen pressure was increased to $1~\mathrm{bar}$ at $700~\mathrm{^{\circ}C}$ before the temperature was lowered to $450~\mathrm{^{\circ}C}$, where the films were annealed \emph{in-situ} for one hour. After removal from the growth chamber every film was annealed \emph{ex-situ} at $450~\mathrm{^{\circ}C}$ for ten hours in flowing oxygen.
The single-layer (SL), trilayer (TL), and bilayer (BL) thin films used for this study are listed in Tab.~\ref{tab:PBCO-studied-films}. The gold capping layers of the first three samples serve for surface-passivation and protection purposes. The single-layer films SL$_{\mathrm{Y}}$ and SL$_{\mathrm{Pr}}$ are mainly used as reference samples for the heterostructures. In order to enhance the signal to background ratio for each of the samples the \lem experiments have been conducted with four films of nominally the same composition.\\
The $c$-axis orientation of the films has been confirmed by X-ray-diffraction measurements. The thicknesses of the layers (cf. Tab.~\ref{tab:PBCO-studied-films}) have been determined within $\sim 3~\mathrm{nm}$ for the MS films and within $\sim 5~\mathrm{nm}$ for the PLD films by Rutherford backscattering spectrometry\cite{Chu-RBS} at the PSI/ETH Laboratory for Ion-Beam Physics. The rms surface roughness of the films measured by atomic-force microscopy is typically $\lesssim 5~\mathrm{nm}$. The critical temperatures of the films obtained by in-plane-resistivity measurements are $T_{\mathrm c} \approx 86~\mathrm{K}$ for SL$_{\mathrm{Y}}$ as well as for BL$_{\mathrm{Y/Pr}}$ and $T_{\mathrm c} \approx 88~\mathrm{K}$ for TL and BL$_{\mathrm{Pr/Y}}$. The width of the transitions and the scattering among nominally identical samples is $\lesssim 2~\mathrm{K}$. As shown in Fig.~\ref{fig:PBCOresistivity} the SL$_{\mathrm{Pr}}$ samples show no superconducting transition but the typical\cite{Fisher-PhysicaC-1991} two-dimensional (2D) VRH in-plane resistivity between $5~\mathrm{K}$ and $100~\mathrm{K}$: $\rho(T)=\rho_0\exp\left[\left(T_0/T\right)^{1/3}\right]$, where $T_0 = 11097~\mathrm{K}$ and $\rho_0 = 4.64\times10^{-4}~\mathrm{\Omega\,cm}$.

\section{Experimental technique}
The \lem experiments have been performed at the $\mu$E$4$ beam line at PSI.\cite{Prokscha-NIMA-2008} Spin-polarized positively charged muons with a tuneable kinetic energy up to $30~\mathrm{keV}$ are implanted into a sample and thermalize. The spins interact with their local environment until the muons decay ($\tau_{\mu}=2.197~\mathrm{\mu s}$)\cite{Nakamura-JPhysG-2010} and emit the decay positrons preferentially in the direction of the muon spin at the time of decay. Thus, by detecting the muons at their implantation time and the positrons after the decay the temporal evolution of the muon-spin polarization (proportional to the decay asymmetry) in a sample can be measured to obtain information about the local environment of the muons. For instance in a static local magnetic field $B_{\mathrm{loc}}$ with a non-zero component perpendicular to the spins $\mathbf{S}_{\mu}$ the muon spins undergo a Larmor precession with a frequency $\omega = \gamma_{\mu} B_{\mathrm{loc}}$, where $\gamma_{\mu} = 2\pi\times 135.54~\mathrm{MHz/T}$ is the gyromagnetic ratio of the muon. More information on $\mu$SR techniques in general and \lem in particular can be found in Refs.~\onlinecite{YaouancReotier} and~\onlinecite{Bakule-ContemporaryPhysics-2004}, respectively. \lem allows a depth-resolved determination of local magnetic fields and field distributions beneath surfaces as well as in thin films and the different layers of heterostructures.\cite{Morenzoni-JPhysCondMatter-2004, Morenzoni-PhysicaB-2009}

\section{Magnetic properties --- zero-field measurements}
\begin{figure}
\centering
\includegraphics[width=0.9\columnwidth]{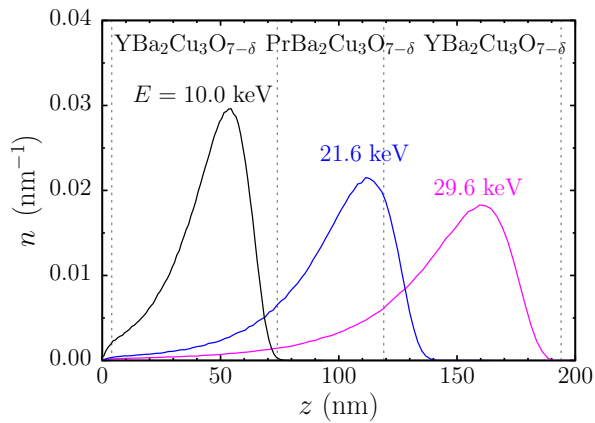}
\caption{(Color online) Selected normalized muon implantation profiles for the TL heterostructure obtained by \trimsp simulations. The dashed lines indicate the boundaries between the different layers of the film. The leftmost and rightmost interfaces represent the boundaries of the heterostructure to the Au top layer and the \STO substrate, respectively.}
\label{fig:PBCO-TL-TRIMSP}
\end{figure}

\begin{figure}
\centering
\includegraphics[width=0.9\columnwidth]{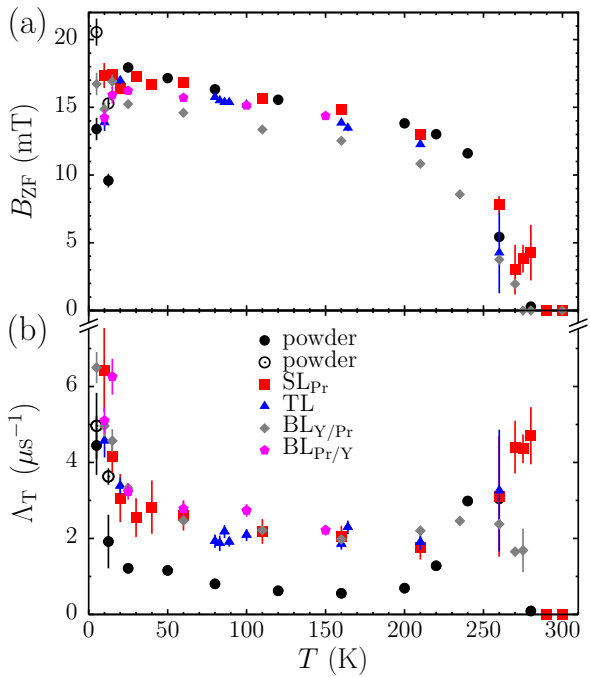}
\caption{(Color online) Temperature dependence of (a) the measured local fields at the muon site and (b) the depolarization rate of the oscillating part of the signal in the heterostructures (cf. Tab.~\ref{tab:PBCO-ZF-energies}) as determined by fits of Eq.~(\ref{eq:PBCO-asy-films}) to the zero-field \lem data. For comparison also the data for a PBCO powder sample from Ref.~\onlinecite{Wojek-PhysicaB-2009} are plotted---there below $T_{\mathrm{N,Pr}}\approx 18$~K a splitting in two frequencies (open and closed symbols) has been observed which cannot be resolved in the thin films.}
\label{fig:PBCOfilms-BandTrate-vs-T}
\end{figure}

In order to characterize the magnetic state of the PBCO layers, zero-field (ZF) \lem experiments have been performed. In each case the muon implantation energies have been chosen adequately, so that according to \trimsp simulations\cite{Eckstein-CompSim-1991} most of the particles stop in the PBCO layer of each heterostructure. The used energies and corresponding stopping fractions are summarized in Tab.~\ref{tab:PBCO-ZF-energies}. Selected muon stopping profiles for the trilayer structure are depicted in Fig.~\ref{fig:PBCO-TL-TRIMSP}.
\begin{table*}
\caption{Fractions of muons stopping in PBCO, in the other layers or being backscattered for different zero-field measurements on the thin-film heterostructures}
\label{tab:PBCO-ZF-energies}
\begin{tabularx}{\linewidth}{C{0.11\linewidth} C{0.09\linewidth} C{0.12\linewidth} C{0.11\linewidth} C{0.20\linewidth} C{0.12\linewidth} C{0.12\linewidth}}
\noalign{\smallskip}
\hline
\noalign{\smallskip}
thin film & sample holder & $E$~(keV) & $\left\langle z_{\mu}\right\rangle$~(nm) & $f_{\mathrm{PrBa}_{2}\mathrm{Cu}_{3}\mathrm{O}_{7-\delta}}$~(\%) & $f_{\mathrm{other}}$~(\%) & $f_{\mathrm{back}}$~(\%) \tabularnewline
\noalign{\smallskip}
\hline
\noalign{\smallskip}
SL$_{\mathrm{Pr}}$ & Ag & $6.0$ & $29$ & $92.4$ & $1.2$ & $6.4$ \tabularnewline
TL & Ag & $21.6$ & $97$ & $67.5$ & $31.1$ & $1.4$ \tabularnewline
BL$_{\mathrm{Y/Pr}}$ & Ag & $23.5$ & $109$ & $90.2$ & $8.8$ & $1.0$ \tabularnewline
BL$_{\mathrm{Pr/Y}}$ & Ni & $8.0$ & $38$ & $96.6$ & $0.0$ & $3.4$ \tabularnewline
\noalign{\smallskip}
\hline
\end{tabularx}
\end{table*}
\begin{figure}
\centering
\includegraphics[width=\columnwidth]{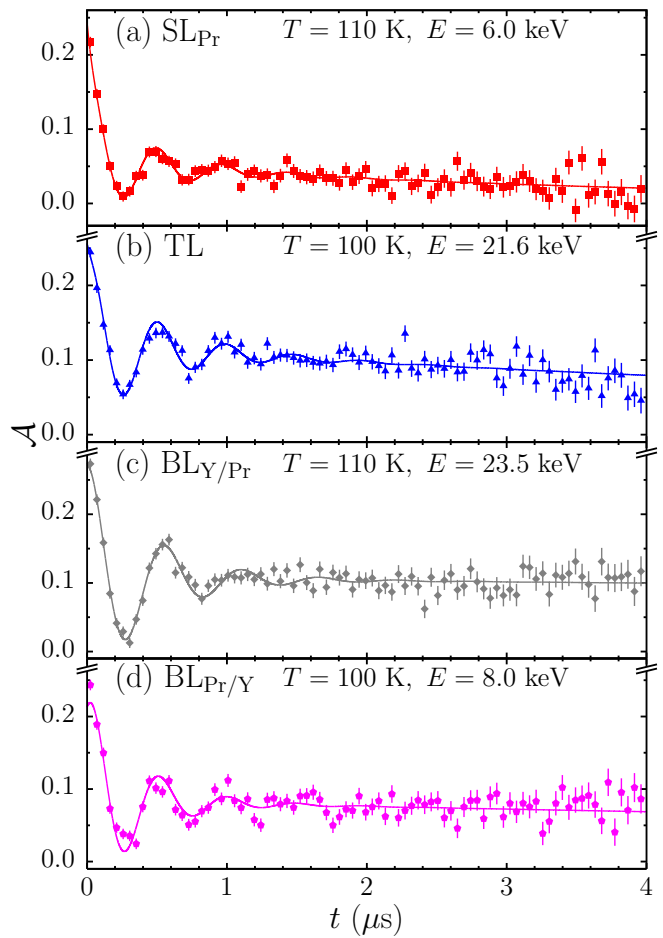}
\caption{(Color online) (a)--(d) Zero-field asymmetry spectra of the studied thin films listed in Tab.~\ref{tab:PBCO-ZF-energies} at $T=100~\mathrm{K}$ or $T=110~\mathrm{K}$, respectively. The muon implantation energies have been chosen to probe mostly the \PBCO layers of the heterostructures. The solid lines are fits to Eq.~(\ref{eq:PBCO-asy-films}).}
\label{fig:PBCOfilms-asy-vs-time}
\end{figure}
Figure~\ref{fig:PBCOfilms-asy-vs-time} shows ZF asymmetry spectra of the different thin films at about $T=100$~K. A spontaneous muon-spin precession caused by the AF order of the planar Cu moments is observed. The Larmor frequency is close to that observed in powder samples.\cite{Wojek-PhysicaB-2009} The larger damping of the precession signal [cf. Fig.~\ref{fig:PBCOfilms-BandTrate-vs-T}(b)] and the fast depolarization at early times (SL$_{\mathrm{Pr}}$ and TL) reflect a higher degree of disorder and strain in the films as compared to the powder; this leads to an overall broader field distribution at the muon site. To analyze the data quantitatively it has to be taken into account that the films are grown with the $c$ axis oriented perpendicular to the large substrate face. Therefore, the magnetic fields detected by the muons are \emph{not} randomly oriented as in the case of the powder samples but mainly point in the direction perpendicular to the film surface and to the muon spins.\cite{Wojek-PhysicaB-2009} Hence, the model function fitted to the data is the following:
\begin{equation}
\begin{split}
\mathcal{A}\left(t\right) = & A\,\Big[ a_{\mathrm{mag,}1} \cos\left(\gamma_{\mu}B_{\mathrm{ZF}}t+\varphi \right) \exp\left(-\Lambda_{\mathrm{T}}t \right) \\ & +  a_{\mathrm{mag,}0} \exp\left(-\Lambda_{0}t \right) \\ & +
\left(1 - a_{\mathrm{mag,}1} - a_{\mathrm{mag,}0}\right) \exp\left(-{\Lambda_{\mathrm{D}}t} \right) \Big].
\end{split}
\label{eq:PBCO-asy-films}
\end{equation}
The first two terms account for the oscillating and the fast depolarizing parts ($10~\mu\mathrm{s}^{-1} \lesssim \Lambda_0 \lesssim 50~\mu\mathrm{s}^{-1}$) of the signal and reflect ordered ($\propto a_{\mathrm{mag},1}$) and more disordered ($\propto a_{\mathrm{mag},0}$) magnetic regions of the PBCO layers, respectively. $\varphi$ represents the virtually temperature-independent initial negative phase of the precession that is often observed in antiferromagnetic cuprates.\cite{Pomjakushin-HyperfineInteract-1994} The last term of Eq.~(\ref{eq:PBCO-asy-films}) arises from contributions of muons stopping in the non-magnetic layers of the heterostructures, within non-magnetic parts of the PBCO, or outside the films in the Ag-coated sample holder. The corresponding depolarization rates are virtually temperature-independent and for all films $\Lambda_{\mathrm{D}}\lesssim 0.3~\mu\mathrm{s}^{-1}$. The total asymmetry $A$ has been fixed to a value of $0.275$, typical for the \lem spectrometer at high implantation energies.
The temperature dependences of the so determined local field $B_{\mathrm{ZF}}$ and depolarization rate $\Lambda_{\mathrm{T}}$ are presented in Fig.~\ref{fig:PBCOfilms-BandTrate-vs-T}. The overall behavior is similar for all the films. However, at intermediate temperatures $\Lambda_{\mathrm{T}}$ is larger by about a factor of $3$ compared to the powder sample. The maxima in $\Lambda_{\mathrm{T}}$ correspond to the antiferromagnetic ordering of the planar Cu moments ($T_{\mathrm{N,Cu}}\approx 280$~K) and the Pr moments ($T_{\mathrm{N,Pr}}\approx18$~K). Also, while the N\'{e}el temperatures are similar in all studied samples and agree well with literature data,\cite{Cooke-PRB-1990, Dawson-HyperfineInteract-1991, Kebede-PRB-1989} the local fields are slightly smaller in the PBCO layers within the heterostructures than in the powder and single layer. This probably reflects structural differences between the various samples which appear most pronounced for BL$_{\mathrm{Y/Pr}}$, where the fields are about $10$~\% to $20$~\% smaller.\\
The magnetic volume fractions of each of the PBCO layers can be estimated by (i) correcting the fractions $a_{\mathrm{mag,}0}$ and $a_{\mathrm{mag,}1}$ of Eq.~(\ref{eq:PBCO-asy-films}) for the part of the muons not stopping in PBCO and (ii) by taking into account instrumental effects, such as the amount of muons backscattered from the sample and not contributing to the signal (especially at the low implantation energies used to study SL$_{\mathrm{Pr}}$ and BL$_{\mathrm{Pr/Y}}$). Considering further a total fraction of about $15$~\% of muons hitting the sample plate, we obtain the following estimates for the magnetic volume fractions within the PBCO layers: $1.0$ (SL$_{\mathrm{Pr}}$), $0.86$ (TL), $0.78$ (BL$_{\mathrm{Y/Pr}}$), and $0.63$ (BL$_{\mathrm{Pr/Y}}$). For the last two samples these numbers are lower limits since only the oscillating part of the asymmetry was used for this estimate ($a_{\mathrm{mag,}0} = 0$). In the first two samples the observed fast depolarizing signal ($a_{\mathrm{mag,}0} \neq 0$) is related to disordered magnetic regions of the samples and therefore contributes to the magnetic volume fraction as well; this may lead to an overestimation.

In summary, the zero-field \lem experiments show that all the $45$~nm to $70$~nm thick PBCO layers of the studied heterostructures display the same antiferromagnetic order as in the bulk. At the same time all the heterostructures contain smaller regions without static magnetic order within the PBCO. While the N\'{e}el temperatures are similar for all specimens the observed local magnetic fields are slightly smaller in the heterostructures and the depolarization rate is considerably enhanced compared to the powder reference sample reflecting a more strongly disordered state and possibly small structural differences in the films.

\section{Superconducting properties --- transverse-field measurements}
The superconducting properties of the thin films have been investigated by means of transverse-field (TF) \lem experiments: After zero-field cooling the samples a small magnetic field up to $\mu_0H=26~\mathrm{mT}$ is applied parallel to the films ($\mathbf{H}\perp\mathbf{c}$, $\mathbf{H}\perp\mathbf{S}_{\mu}$). In this geometry, the screening of an external magnetic field within a superconductor in the Meissner state can be observed directly by measuring the local magnetic fields as a function of the muon implantation depth.\cite{Jackson-PRL-2000, Suter-PRB-2005, Kiefl-PRB-2010}\\
In the TF measurements of the heterostructures there are three contributions to the \lem signals: The muons implanted into the AF regions of the PBCO layer probe a superposition of the internal and the applied fields.\cite{Wojek-PhysicaB-2009} The muons stopping in the YBCO layers sense a paramagnetic (above \Tc) or diamagnetic (below \Tc) environment. The fraction of muons thermalizing in regions of PBCO without static order contributes to this signal as well. Finally, a ``background signal'' arises from muons missing the samples and stopping in the {Ag}-coated sample plate where their spins precess at a frequency corresponding to the applied field. Since the absolute diamagnetic field shift in these thin films is small, the field distributions probed by the muons are rather narrow and can be well approximated by a Gaussian distribution so that the \lem asymmetry signal can be analyzed using the following function
\begin{equation}
\begin{split}
\mathcal{A}(t) & = A_{\mathrm{mag}} \cos\left(\gamma_{\mu}B_{\mathrm{mag}}t+\varphi_{\mathrm{mag}}\right) \exp\left(-\Lambda t\right) \\ & + A_{\mathrm{sc}} \cos\left(\gamma_{\mu}B_{\mathrm{sc}}t+\varphi\right) \exp\left(-\frac{1}{2}\sigma_{\mathrm{sc}}^2t^2\right) \\ & + A_{\mathrm{bg}} \cos\left(\gamma_{\mu}B_{\mathrm{bg}}t+\varphi\right) \exp\left(-\frac{1}{2}\sigma_{\mathrm{bg}}^2t^2\right).
\end{split}
\label{eq:PBCO-TL-TFasymmetry}
\end{equation}
Here, $\varphi_{\mathrm{mag}}$ and $\varphi$ are the initial phases of the respective precession signals. They are distinct due to the different axes of precession given by the corresponding local fields. The background fraction $A_{\mathrm{bg}}$ as a function of the muon implantation energy and the applied field has been estimated by independent TF measurements of a {Ag} foil of the same dimensions as the heterostructures mounted on a Ni-coated sample holder. Here muons stopping outside the {Ag} sample in the ferromagnetic {Ni} depolarize quickly whereas the spins of the muons in the {Ag} precess in the applied field. Thus, the muon fraction in {Ni} can be used as estimate for $A_{\mathrm{bg}}$ and this parameter is kept fixed during the analysis of the film data. The average local field $B_{\mathrm{mag}}$ within the AF regions and the corresponding spin-depolarization rate $\Lambda$ have been determined at the implantation energy where most of the muons thermalize in the PBCO layers. At the other energies these parameters have been fixed to reduce correlations between different fit parameters.
\begin{figure}
\centering
\includegraphics[width=\columnwidth]{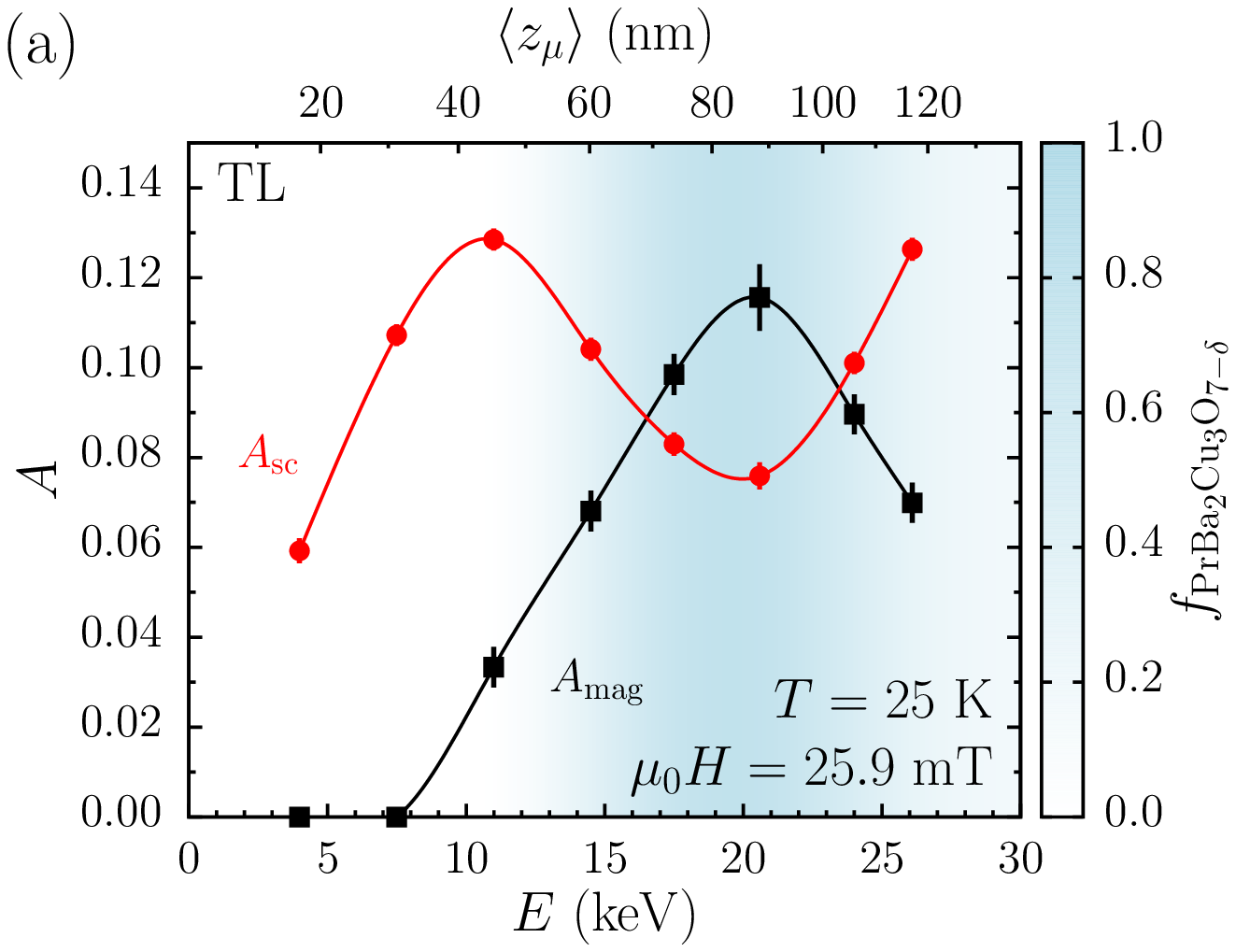}
\includegraphics[width=\columnwidth]{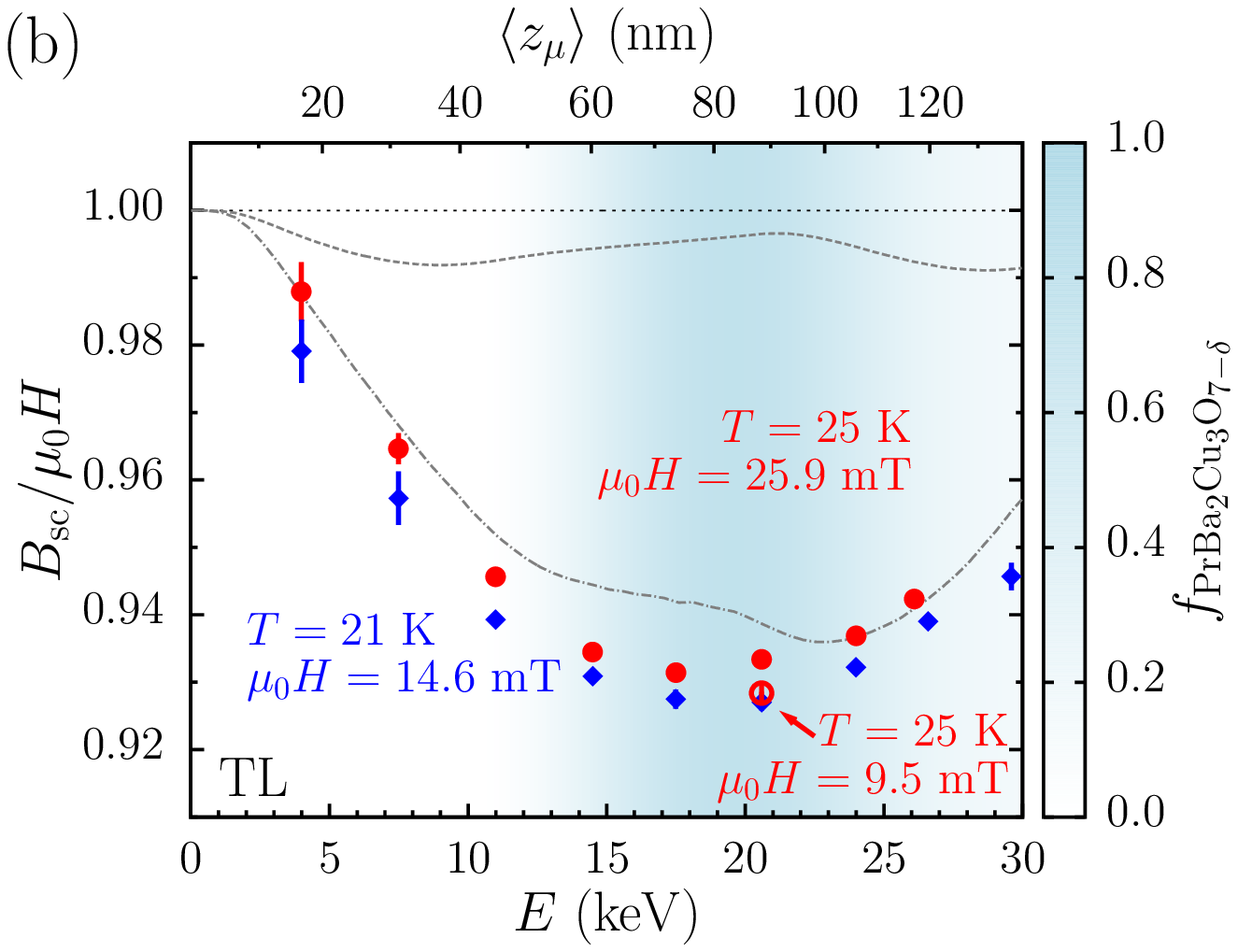}
\caption{(Color online) (a) Measured partial asymmetries $A_{\mathrm{mag}}$ and $A_{\mathrm{sc}}$ in the TL sample as a function of the muon implantation energy $E$ (lower scale) and the mean muon implantation depth $\langle z_{\mu}\rangle$ (upper scale) at $T=25~\mathrm{K}$ and in an applied field of $25.9~\mathrm{mT}$. The missing part of the total asymmetry of about $0.27$ at high and at least $0.20$ at the lowest energies is mainly contained in the background $A_{\mathrm{bg}}$. The solid lines are guides to the eye. (b) Muon-energy dependence of the normalized diamagnetically shifted local fields at various applied fields at $T_{\mathrm{N,Pr}}<T\ll T_{\mathrm{c}}$. The dashed and the dash-dotted lines represent the expected field screening in the TL heterostructure in case of a fully insulating barrier and a barrier containing shorts in $c$ direction, respectively. These curves are obtained using a local London model introduced in the text ($\lambda_{\mathrm{eff}}^{\mathrm{YBCO}}=210~\mathrm{nm}$). In both graphs the background shading indicates the fraction of the implanted muons stopping in the PBCO barrier layer.}
\label{fig:TLscreening}
\end{figure}
For an applied field of $25.9~\mathrm{mT}$ the energy dependences of the asymmetries $A_{\mathrm{mag}}$ and $A_{\mathrm{sc}}$ are shown in Fig.~\ref{fig:TLscreening}(a). The AF order in the PBCO barrier as observed in the zero-field experiments is hardly disturbed and the magnetic volume fraction of PBCO determined from the partial asymmetries in the TF measurements is found to be $\approx 90$~\% in the films. In Fig.~\ref{fig:TLscreening}(b) the screened fields $B_{\mathrm{sc}}$ as a function of the muon implantation energy in the TL sample are depicted for various applied fields and temperatures of $21~\mathrm{K}$ and $25~\mathrm{K}$, respectively. These temperatures are sufficiently below \Tc of the YBCO layers but above $T_{\mathrm{N,Pr}}$, where the Pr ordering strongly broadens the signal from the PBCO layers. The experiment shows unexpected field screening throughout the whole heterostructure and the field profile is similar to the hyperbolic-cosine field penetration into a homogeneous thin superconducting film. For the TL system in the absence of Cooper-pair transport across the PBCO barrier one would expect to find only modest screening of the applied field in the top and bottom YBCO layers [cf. dashed line in Fig.~\ref{fig:TLscreening}(b)].
\begin{figure}
\centering
\includegraphics[width=\columnwidth]{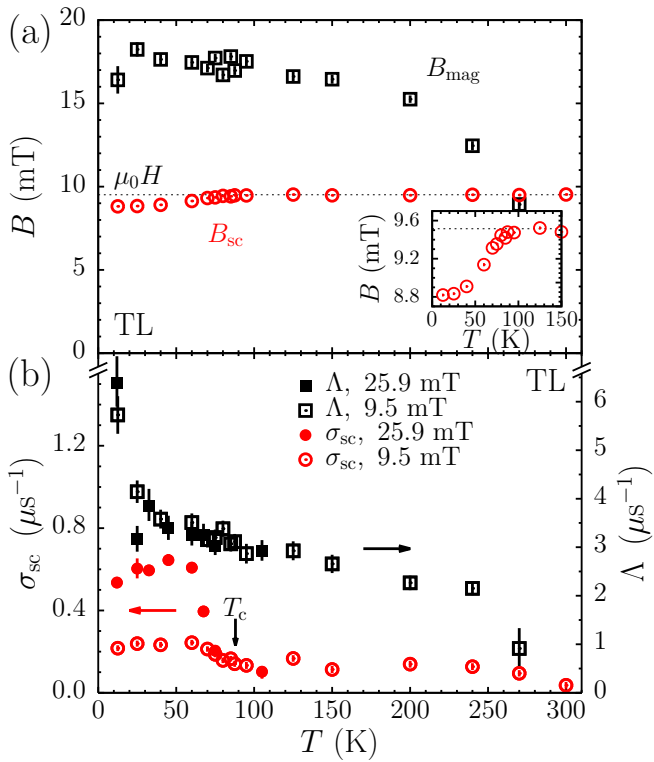}
\caption{(Color online) (a) Temperature dependence of the local magnetic fields $B_{\mathrm{sc}}$ and $B_{\mathrm{mag}}$ measured in an applied field of $9.5~\mathrm{mT}$ probing the center of the TL structure using a muon implantation energy of $20.6~\mathrm{keV}$. (b) Depolarization rates $\sigma_{\mathrm{sc}}$ and $\Lambda$ at the same energy for the highest and lowest applied fields as a function of temperature.}
\label{fig:TLfields-vs-T}
\end{figure}
The temperature dependence of the observed local magnetic fields in the center of the trilayer is shown in Fig.~\ref{fig:TLfields-vs-T}(a). Clearly, the screening of the applied field in the structure sets in below \Tc of YBCO. Since the applied and the internal fields in PBCO [$B_{\mathrm{ZF}}$, cf. Fig.~\ref{fig:PBCOfilms-BandTrate-vs-T}(a)] are roughly perpendicular to each other $\displaystyle B_{\mathrm{mag}}^2 \approx {B_{\mathrm{ZF}}^2 + \left(\mu_0H\right)^2} \approx {B_{\mathrm{ZF}}^2 + B_{\mathrm{sc}}^2}$. Given the uncertainty in the determination of $B_{\mathrm{mag}}$ it cannot be concluded whether the applied or the diamagnetically shifted field contributes to the signal in the AF regions.
As shown in Fig.~\ref{fig:TLfields-vs-T}(b) for the applied fields of $9.5$~mT and $25.9$~mT, parallel to the decrease of $B_{\mathrm{sc}}$ below \Tc the corresponding muon-spin-depolarization rate $\sigma_{\mathrm{sc}}$ increases, thus reflecting the broadened field distribution due to the partial screening of the applied field.
The temperature dependence of $\Lambda$ reflects---as in the ZF case [Fig.~\ref{fig:PBCOfilms-BandTrate-vs-T}(b)]---the ordering of the Cu moments at $T_{\mathrm{N,Cu}}\approx 280$~K and that of the Pr moments at $T_{\mathrm{N,Pr}}\approx18$~K.

The diamagnetism observed in the trilayer structure indicates that supercurrents can enter and cross the PBCO barrier even though a single PBCO layer is overall not superconducting. In principle in a TL structure the possibility of superconducting shorts connecting the two ``electrodes'' has to be considered [cf. dash-dotted line in Fig.~\ref{fig:TLscreening}(b)]. In the past such shorts have been found to affect results of HTS Josephson devices.\cite{Delin-SupercondSciTechnol-1996}
Yet, for the barrier layer with the relatively large thickness of $45$~nm the probability of occurring superconducting filaments is rather small. Moreover, a rough estimate shows that even if present they would be driven normal by the supercurrents associated with the field screening. To reduce the possibility of outgrown shorts we repeated the \lem experiments after etching away the edges (about $0.5~\mathrm{mm}$ on each side) of the TL films which are mostly at risk to contain superconducting shorts. The very same results were obtained---the previously shown data with an applied field of $14.6~\mathrm{mT}$ have been taken on the etched films [cf. Fig.~\ref{fig:TLscreening}(b)].
However, the most convincing demonstration that the effect (Meissner screening in PBCO due to the proximity to YBCO) is intrinsic is given by measurements on the bilayer structures $\mathrm{BL}_{\mathrm{Y/Pr}}$ and $\mathrm{BL}_{\mathrm{Pr/Y}}$ discussed in the following. With only one superconducting layer present, the occurrence of microshorts and the related spurious effects are excluded on principle.
\begin{figure}
\centering
\includegraphics[width=\columnwidth]{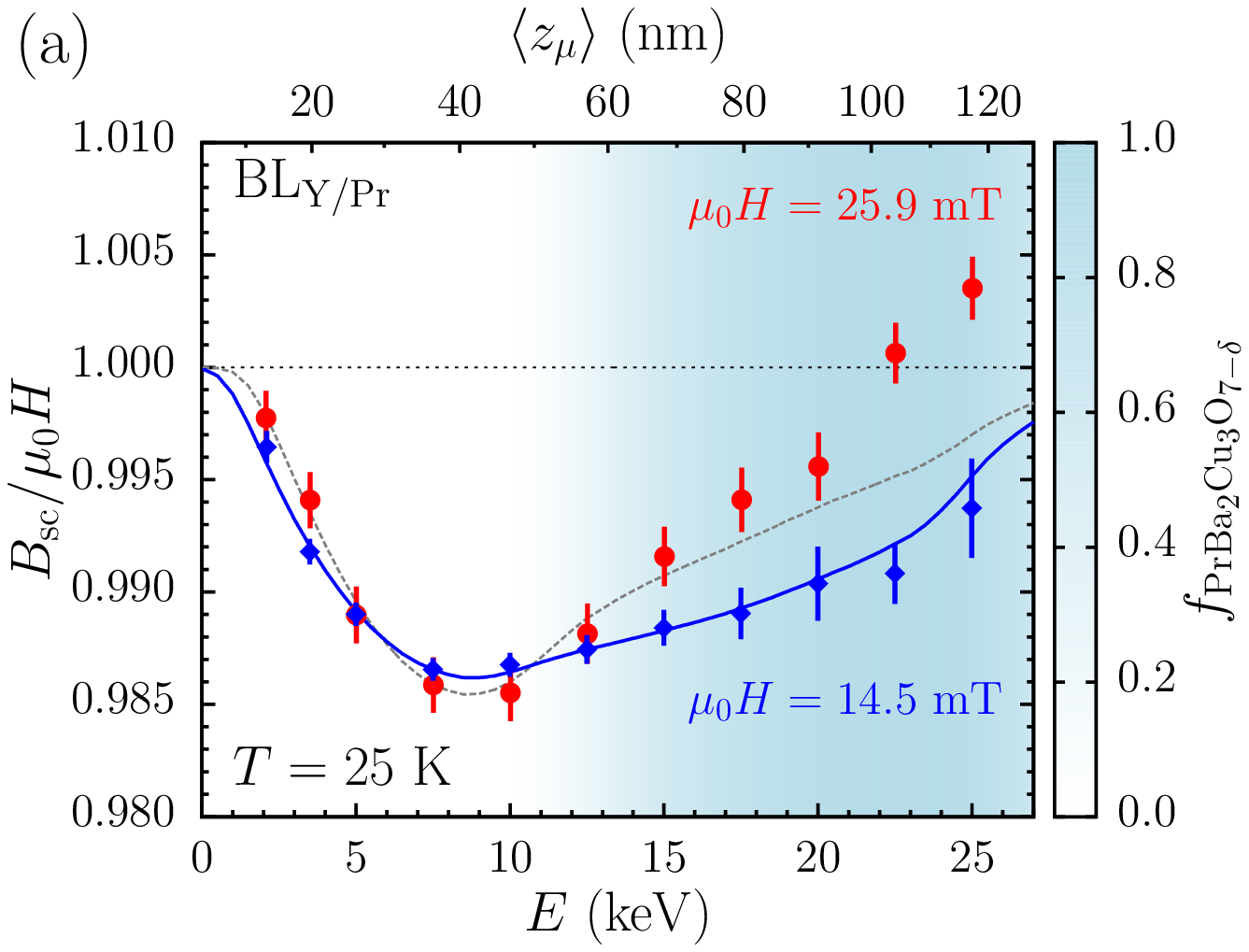}
\includegraphics[width=\columnwidth]{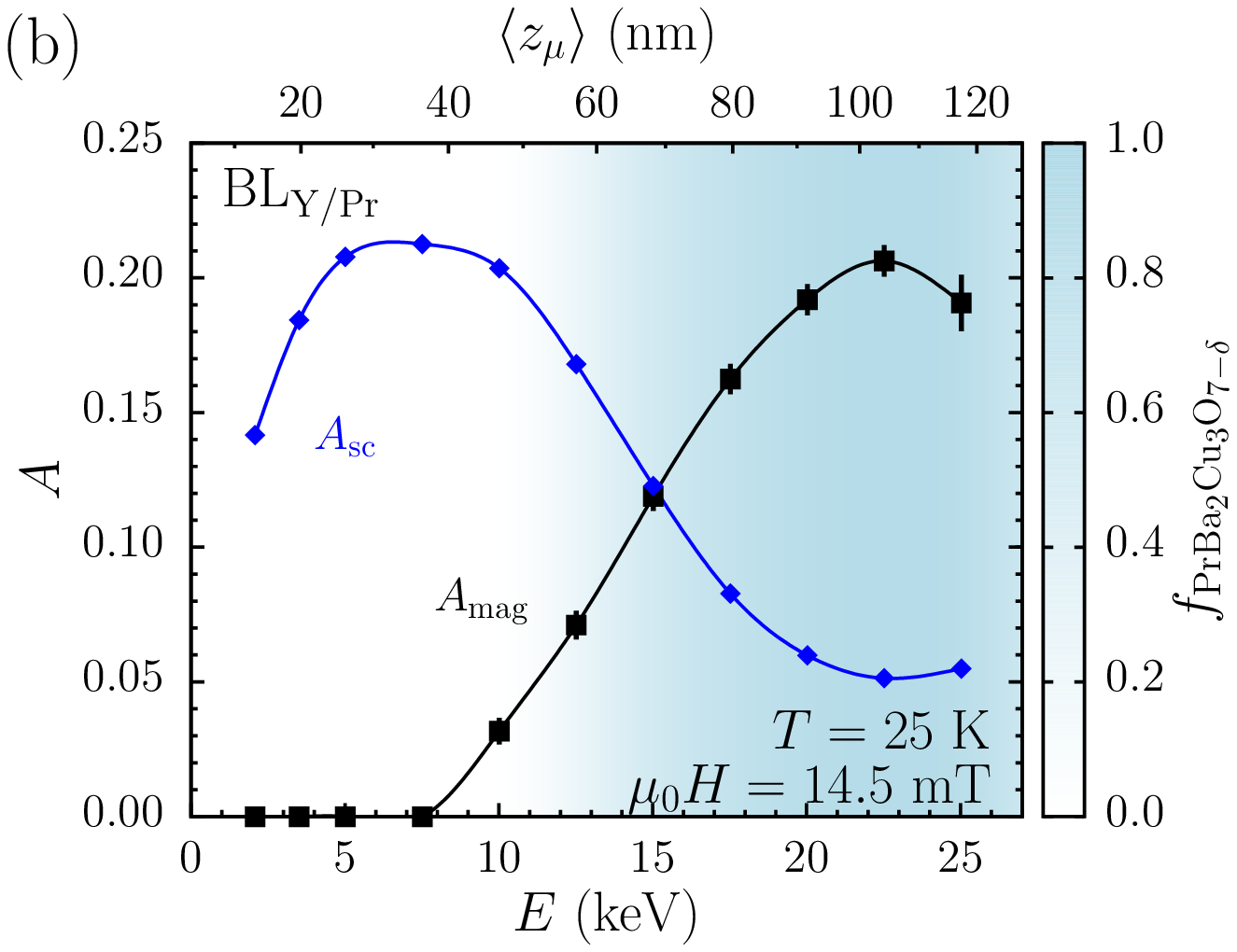}
\caption{(Color online) (a) Normalized screened field as a function of the muon implantation energy in $\mathrm{BL}_{\mathrm{Y/Pr}}$ at $T=25~\mathrm{K}$ for two applied fields of $14.5~\mathrm{mT}$ and $25.9~\mathrm{mT}$. The solid and dashed lines represent fits of the phenomenological London model introduced in the text. (b) Energy dependence of the partial asymmetries $A_{\mathrm{sc}}$ and $A_{\mathrm{mag}}$. The lines are guides to the eye. The background shading indicates the fraction of the implanted muons stopping in the PBCO layer.}
\label{fig:BLscreening}
\end{figure}

As shown in Fig.~\ref{fig:BLscreening}(a) below \Tc of YBCO a small but sizeable diamagnetic shift of the applied magnetic field is observed also in the $\mathrm{BL}_{\mathrm{Y/Pr}}$ sample at energies where most of the muons probe the PBCO layer. The corresponding partial asymmetries are presented in Fig.~\ref{fig:BLscreening}(b). Since at all energies a fraction of muons stops in the superconducting YBCO top layer, to detect a possibly present diamagnetic signal arising from muons implanted in the AF PBCO layer a careful analysis is needed. For this, quantitative information on the background signal of muons missing the sample is required. This has been obtained by two additional measurements: Firstly, at $T=300~\mathrm{K}>T_{\mathrm{N,Cu}}$ an energy scan has been performed to determine accurately the background field for the different muon energies. Secondly, as mentioned above, measurements of a {Ag} foil with the same dimensions as the thin-film mosaic mounted on a {Ni}-coated sample holder were used to determine the sample and background contributions to the total \lem signal. For the analysis of the low-temperature data with Eq.~(\ref{eq:PBCO-TL-TFasymmetry}) the background contribution and the values of $B_{\mathrm{mag}}$ and $\Lambda$---determined at the highest implantation energy---have been fixed.
The results in Fig.~\ref{fig:BLscreening}(a) show that at least at the low applied field the obtained field profile can only be explained by the presence of supercurrents also in the PBCO layer. To quantify the superconducting properties of the $\mathrm{BL}_{\mathrm{Y/Pr}}$ sample we applied a phenomenological one-dimensional London model. To calculate the magnetic induction throughout the structure, we model a juxtaposition of two local superconductors with different effective in-plane magnetic-field penetration depths $\lambda_{\mathrm{eff}}$:
\begin{equation}
B_i(z) = \tilde{a}_i \exp\left(-z/\lambda_{\mathrm{eff},\,i}\right) + \tilde{b}_i \exp\left(z/\lambda_{\mathrm{eff},\,i}\right).
\label{eq:London-model-for-two-sc}
\end{equation}
Here $i=1,2$ and the coefficients $\tilde{a}_i$ as well as $\tilde{b}_i$ are determined by the boundary conditions:\cite{London-BoundaryConditions}
\begin{equation}
\begin{split}
B_1(0) & = B_2(d) = \mu_0H \\
B_1(d') & = B_2(d') \\
\lambda_{\mathrm{eff},1}^2 \frac{\partial B_1(z)}{\partial z}\Bigg\vert_{z=d'} & = \lambda_{\mathrm{eff},2}^2 \frac{\partial B_2(z)}{\partial z}\Bigg\vert_{z=d'},
\end{split}
\label{eq:London-model-for-two-sc-bc}
\end{equation}
where $z=0$ and $z=d$ denote the vacuum and the substrate interface and $z=d'$ is the interface between the different layers. An additional non-screening ``dead layer'' with a thickness of $5~\mathrm{nm}$ to $10~\mathrm{nm}$ at the surface mostly caused by surface-roughness effects has to be taken into account. The induction then only decays below this dead layer and not from the vacuum interface. Using the so-defined field penetration into the thin film, the calculated muon implantation profiles and the fact that only a small part of the muons stopping in the PBCO layer contributes to $A_{\mathrm{sc}}$, the data can be analyzed using this model for the mean-field values in order to obtain $\lambda_{\mathrm{eff},i}$. Two cases are shown in Fig.~\ref{fig:BLscreening}(a): The solid blue line takes into account two finite penetration depths $\lambda_{\mathrm{eff}}^{\mathrm{YBCO}}=210(20)~\mathrm{nm}$ and $\lambda_{\mathrm{eff}}^{\mathrm{PBCO}}=740(50)~\mathrm{nm}$, the dashed grey line allows only the top layer to be superconducting with $\lambda_{\mathrm{eff}}^{\mathrm{YBCO}}=160(20)~\mathrm{nm}$, whereas $B=\mu_0H$ in the non-magnetic regions of the bottom layer. The comparison with the data shows that the diamagnetic response in an applied field of $14.5~\mathrm{mT}$ can only be consistently explained by a finite induced superfluid density in the PBCO layer. By contrast, in the increased field of $25.9~\mathrm{mT}$ the effective screening in the bottom layer is negligible and only the region within $5~\mathrm{nm}$ from the interface contributes with fields smaller than the applied field. For higher implantation depths the measured fields even exceed the applied field; this might be due to flux expelled from the superconducting part of the sample enhancing the induction in the neighboring layer.
These observations are consistent with the overall findings in the trilayer: There seems to be a substantial proximity effect inducing superfluid density in thick and otherwise non-superconducting PBCO layers. Also, the suppression of this effect in higher magnetic fields is apparent in the TL data [Fig.~\ref{fig:TLscreening}(b)]. Employing the phenomenological London model to the TL system yields an effective penetration depth for YBCO of the same order of magnitude as for the bilayer. However, since the top and bottom layers dominate the diamagnetic response of the TL structure, an accurate determination of the effective penetration depth in the barrier is not possible from the TL data.

\begin{figure}
\centering
\includegraphics[width=\columnwidth]{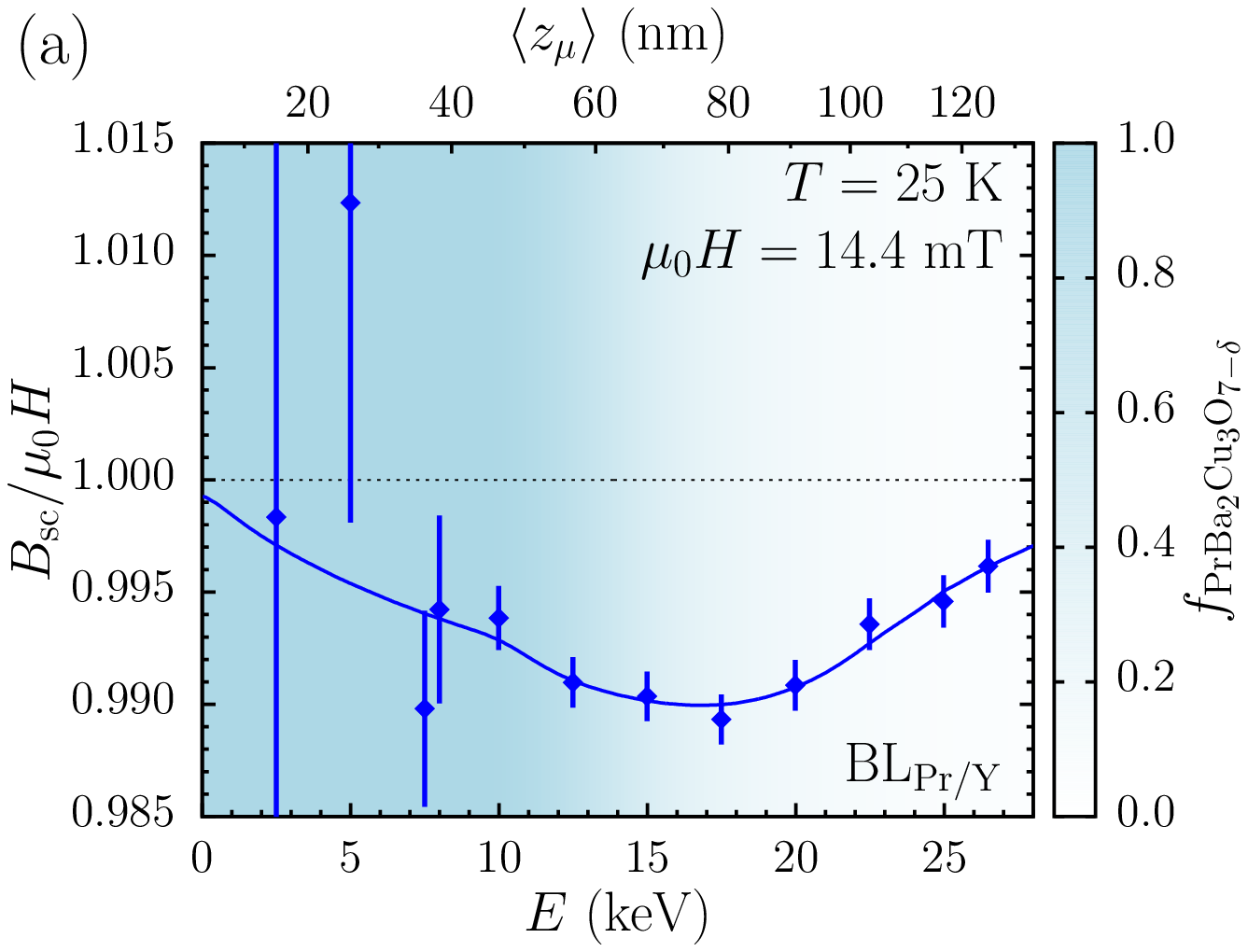}
\includegraphics[width=\columnwidth]{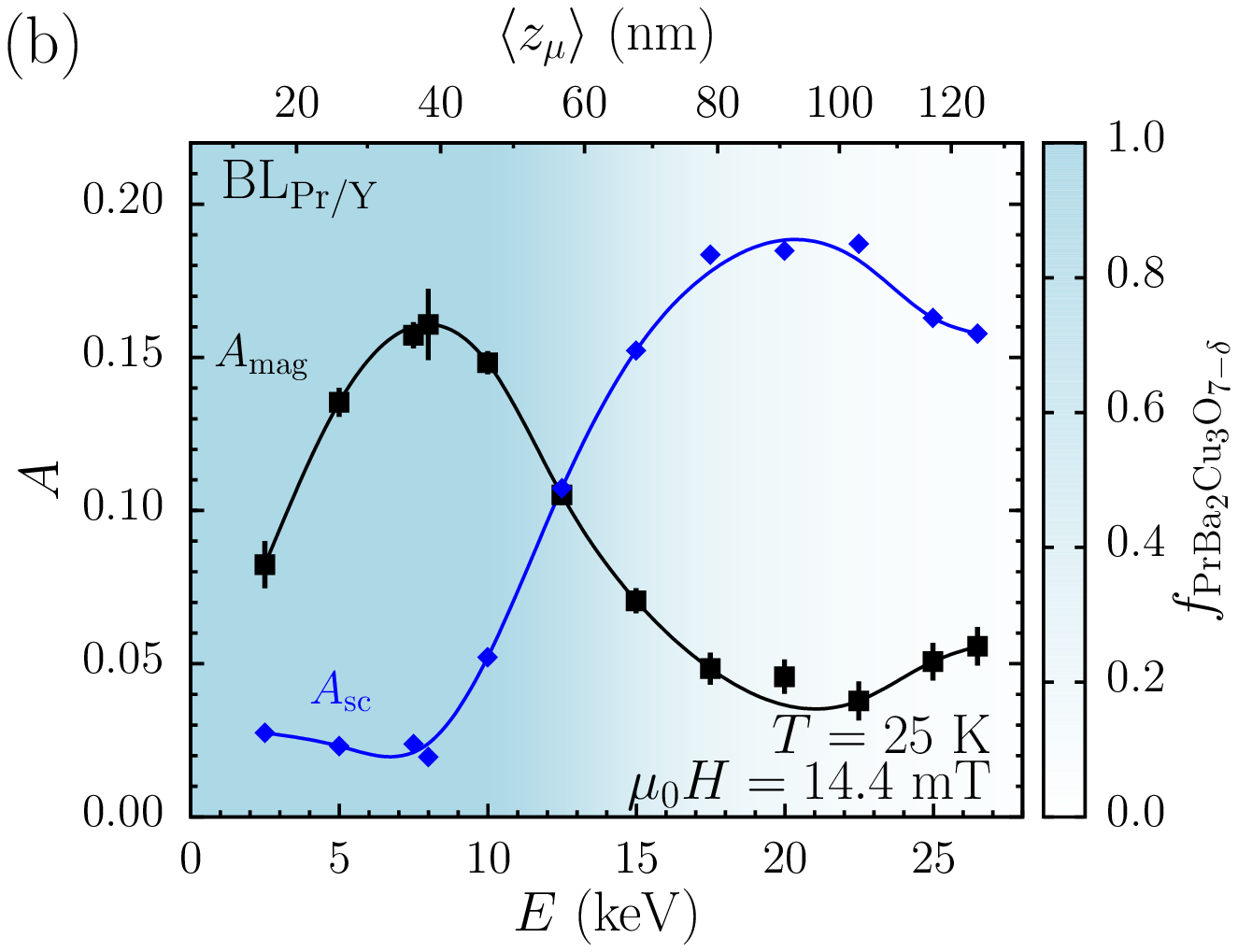}
\caption{(Color online) (a) Normalized screened field as a function of the muon implantation energy in $\mathrm{BL}_{\mathrm{Pr/Y}}$ at $T=25~\mathrm{K}$ for an applied field of $14.4~\mathrm{mT}$. The solid line represents a fit to the phenomenological London model introduced in the text yielding $\lambda_{\mathrm{eff}}^{\mathrm{YBCO}}=315(30)~\mathrm{nm}$ and $\lambda_{\mathrm{eff}}^{\mathrm{PBCO}}=650(50)~\mathrm{nm}$. (b) Energy dependence of the partial asymmetries $A_{\mathrm{sc}}$ and $A_{\mathrm{mag}}$. The lines are guides to the eye. The background shading indicates the fraction of the implanted muons stopping in the PBCO layer.}
\label{fig:PBCO-BLpry-BandAsym-vs-E}
\end{figure}
Finally, TF \lem experiments in the Meissner geometry have also been conducted on the $\mathrm{BL}_{\mathrm{Pr/Y}}$ bilayer. In this case the antiferromagnetic PBCO layer of the structure can be probed at energies where no muons stop in the underlying \YBCO layer, thus eliminating any uncertainty which may arise from the stopping profiles. Here, the mosaic of the four films has been mounted on a Ni-coated sample plate in order to remove background contributions from muons missing the sample, and therefore, $A_{\mathrm{bg}} = 0$ in Eq.~(\ref{eq:PBCO-TL-TFasymmetry}); otherwise the analysis is performed as described above. The screened fields and the asymmetries are shown as a function of the muon implantation energy in Fig.~\ref{fig:PBCO-BLpry-BandAsym-vs-E} for $T=25~\mathrm{K}$ and an applied field of $14.4~\mathrm{mT}$. The energy-independent $A_{\mathrm{sc}}$ in the PBCO layer indicates the presence of spatially inhomogeneous disordered magnetic areas---for a completely homogeneous layer $A_{\mathrm{sc}}$ would initially increase as a function of the implantation energy qualitatively similar to $A_{\mathrm{mag}}$ reflecting the decreasing fraction of backscattered and reflected muons with increasing energy. At $E=8~\mathrm{keV}$ $A_{\mathrm{sc}}$ is about $10$~\% of the total asymmetry (and the magnetic volume fraction in PBCO is about $90$~\%, accordingly). In the zero-field measurements this fraction could not be determined accurately because of the contributions of the rapidly depolarizing Ni signal. Due to the rather small $A_{\mathrm{sc}}$ it is not possible to obtain a precise field profile $B_{\mathrm{sc}}$ at energies where the muons stop solely in the PBCO top layer [Fig.~\ref{fig:PBCO-BLpry-BandAsym-vs-E}(a)]. That indeed a screening of the applied magnetic field is also effective in this case becomes evident from the local fields measured between $E=10~\mathrm{keV}$ and $E=15~\mathrm{keV}$ where the implanted muons partially penetrate the YBCO layer of the structure. If the supercurrents were confined to the bottom layer, the fields in that region should be much closer to the applied field. Employing the London model to describe the measured mean fields for $\mathrm{BL}_{\mathrm{Pr/Y}}$ yields $\lambda_{\mathrm{eff}}^{\mathrm{YBCO}}=315(30)$~nm and $\lambda_{\mathrm{eff}}^{\mathrm{PBCO}}=650(50)$~nm at $T=25~\mathrm{K}$. The value for PBCO is consistent with the result of the other bilayer, although the larger magnetic penetration depth in YBCO might indicate that the $\mathrm{BL}_{\mathrm{Pr/Y}}$ films are more defective or oxygen-deficient than the reversed structures.
\begin{figure}
\centering
\includegraphics[width=0.9\columnwidth]{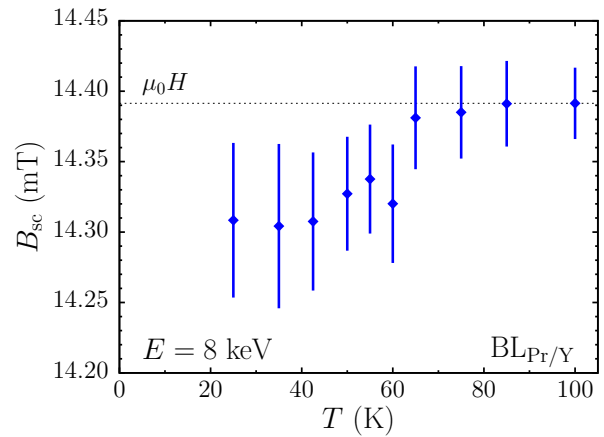}
\caption{Temperature dependence of the local screened field $B_{\mathrm{sc}}$ measured at $E=8~\mathrm{keV}$ in the PBCO top layer of the $\mathrm{BL}_{\mathrm{Pr/Y}}$ heterostructure.}
\label{fig:PBCO-BLpry-B-vs-T}
\end{figure}
Moreover, the induced diamagnetism in the PBCO layer can be followed as a function of temperature. Figure~\ref{fig:PBCO-BLpry-B-vs-T} shows $B_{\mathrm{sc}}$ between $T=25$~K and $T=100$~K for an implantation energy $E=8$~keV, where all muons stop in the PBCO top layer. Although small, a systematic diamagnetic shift below about $60$~K is observed in this case, thus reinforcing the detection of diamagnetism in PBCO adjacent to YBCO.

\section{Discussion}
The observed proximity effect could be related to the intrinsic peculiarities of the PBCO electronic structure. The valence state of the Pr ions in PBCO is mostly Pr$^{3+}$ (like Y$^{3+}$ in YBCO).\cite{Lopez-Morales-PRB-1990, Soderholm-PRB-1991, Staub-PRB-2000} A priori, no reduction of the carrier density and no suppression of superconductivity in this material would be expected compared to other members of the YBCO family of high-temperature superconductors. The most favored model explains the insulating behavior of PBCO with a hybridization between {Pr}-$4f$ and {O}-$2p_{\pi}$ orbitals which leads to an effective hole localization and therefore insulating {CuO$_2$} planes while the {CuO} chains remain metallic locally.\cite{Fehrenbacher-PRL-1993} This model has been further extended\cite{Liechtenstein-PRL-1995, Mazin-PRB-1998} to conclude that the relevant level has a finite bandwidth and the carrier localization might be the result of disorder. Experimentally, the O-$2p_{\pi}$ character of the holes has been confirmed by X-ray absorption spectroscopy.\cite{Merz-PRB-1997, Merz-PRB-1999} However, within the model of Ref.~\onlinecite{Fehrenbacher-PRL-1993} small variations in the occupation of the $4f$ states yield either an insulating or a metallic system and therefore, local structural changes inducing more metallic regions could explain the observed proximity effect.\\

Also, there are reports on inhomogeneous superconductivity in PBCO which is destroyed by {Pr$^{3+}$} magnetic ions on {Ba$^{2+}$} lattice sites,\cite{Blackstead-PRB-1996, Shukla-PRB-1999} bulk superconductivity in single crystals grown by the traveling-solvent floating-zone technique which have longer $c$ axes compared to crystals grown by standard solid-state-reaction techniques,\cite{Zou-PRL-1998} and superconducting clusters in phase-separated {Pr}-rich material.\cite{Grevin-PRB-2000} Additionally, it has been suggested\cite{Mazin-PRB-1999} that PBCO is a novel type of superconductor where the mobile carriers reside in the Pr-O bands---and not in the Cu-O bands as in other cuprate high-temperature superconductors. Yet, recent band-structure calculations have yielded a metallic and possibly superconducting state in PBCO (similar to that of YBCO) which is likely to be destroyed by disorder.\cite{Ghanbarian-PRB-2008} Thus, it is possible that the observed non-magnetic regions within the PBCO layers might be less disordered and therefore generally closer to a local superconducting state in which the long-range coherence is enhanced by the presence of the superconducting YBCO ``electrodes''. This would also provide a natural explanation for the increased interlayer-coupling effects observed in several transport measurements in Josephson junctions and superlattices with PBCO barrier layers.\\

Yet another possible scenario is motivated by the preserved AF order in the CuO$_2$ planes of the PBCO layers. This might indicate that the supercurrent transport is governed by tunneling through localized states in the {CuO} chains while most of the {CuO$_2$} planes remain insulating.

\section{Conclusion}
The observation of the Meissner effect in trilayer \emph{and} bilayer heterostructures demonstrates that superconductivity can be induced in PBCO layers by the proximity to YBCO layers.
The effect occurs in few-dozen-nanometer thick PBCO layers which are non-superconductive but ``semiconducting'' if grown as single-layer films. Furthermore, for low applied magnetic fields an effective penetration depth of the order of $650~\mathrm{nm}$ to $750~\mathrm{nm}$ at $T=25~\mathrm{K}$ can be attributed to the induced superconducting state. However, this effect is considerably decreased already in an applied magnetic field of about $26~\mathrm{mT}$. At the same time, our results show that the intrinsic AF ordering of the planar {Cu} spins in about $90~\mathrm{\%}$ of the volume of the PBCO layers is hardly disturbed by the Cooper-pair transport through the material. The observed effect might indicate either local structural changes yielding partially metallic regions within PBCO or enhanced tunneling through localized states in the {CuO} chains.
To separate the contributions of the possibly involved mechanisms and to gain a better insight in the length scales of the effect, it would be desirable to determine the trilayer field screening as a function of the barrier thickness.

\acknowledgments
The $\mu$SR measurements were performed at the Swiss Muon Source, Paul Scherrer Institut, Villigen, Switzerland. This work has been supported by the Swiss National Science Foundation and the NCCR MaNEP.

\input{manuscript-arXiv-20111213.bbl}

\end{document}

%% file: manuscript-arXiv-20111213.bbl
%